\documentclass[conference,twocolumn]{IEEEtran}
\usepackage{amsmath}
\usepackage{graphicx}
\usepackage{amsthm}
\usepackage{xfrac}
\usepackage{amssymb}
\usepackage{cite}
\usepackage{xcolor,accents}
\usepackage[font=small,skip=4pt]{caption}

\graphicspath{{Figures/}}

\newtheorem{Lemma}{Lemma}

\newtheorem{Theorem}{Theorem}

\DeclareMathSymbol{\widehatsym}{\mathord}{largesymbols}{"62}
\newcommand\lowerwidehatsym{%
  \text{\smash{\raisebox{-1.3ex}{%
    $\widehatsym$}}}}
\newcommand\fixwidehat[1]{%
  \mathchoice
    {\accentset{\displaystyle\lowerwidehatsym}{#1}}
    {\accentset{\textstyle\lowerwidehatsym}{#1}}
    {\accentset{\scriptstyle\lowerwidehatsym}{#1}}
    {\accentset{\scriptscriptstyle\lowerwidehatsym}{#1}}
}

\IEEEoverridecommandlockouts

\begin{document}

\title{Performance of Cell-Free Massive MIMO Systems with MMSE and LSFD Receivers}

\author{\IEEEauthorblockN{Elina Nayebi}
\IEEEauthorblockA{University of California\\
San Diego, CA 92093\thanks{\hrule\vspace{0.4ex}The work of E. Nayebi and B. Rao were supported by the National Science Foundation under Grant CCF-1115645.}}
\and
\IEEEauthorblockN{Alexei Ashikhmin}
\IEEEauthorblockA{Bell Laboratories\\
Murray Hill, NJ 07974}
\and
\IEEEauthorblockN{Thomas L. Marzetta}
\IEEEauthorblockA{Bell Laboratories\\
Murray Hill, NJ 07974}
\and
\IEEEauthorblockN{Bhaskar D. Rao}
\IEEEauthorblockA{University of California\\
San Diego, CA 92093}
}

\maketitle

\IEEEpeerreviewmaketitle

\begin{abstract}
%

Cell-Free Massive MIMO comprises a large number of distributed single-antenna access points (APs) serving a much smaller number of users. There is no partitioning into cells and each user is served by all APs.

In this paper, the uplink performance of cell-free systems with minimum mean squared error (MMSE) and large scale fading decoding (LSFD) receivers is investigated. The main idea of LSFD receiver is to maximize achievable throughput using only large scale fading coefficients between APs and users. Capacity lower bounds for MMSE and LSFD receivers are derived. An asymptotic approximation for signal-to-interference-plus-noise ratio (SINR) of MMSE receiver is derived as a function of large scale fading coefficients only. The obtained
approximation is accurate even for a small number of antennas.
MMSE and LSFD receivers demonstrate five-fold and two-fold gains respectively
over matched filter (MF) receiver in terms of $5\%$-outage rate.


\end{abstract}


\section{Introduction}

In recent years Massive MIMO (mMIMO) has attracted considerable attention as a candidate for the fifth generation physical layer technology \cite{marzetta2010},\cite{rusek2013scaling}. Cell-free mMIMO is a particular deployment of mMIMO systems with a network of randomly-located large number of
single-antenna APs, where the geographical area is not partitioned into cells and each user is served simultaneously by all of the APs \cite{cellfree},\cite{Hien}.

Some of the advantages and limitations of the networks
with distributed APs can be found in \cite{cellfree,Hien,gesbert2010multi,sawahashi2010coordinated,truong2013viability}.
In particular in [3], [4] the performance of downlink transmission and uplink transmission with MF receiver in cell-free mMIMO systems have been studied. In this paper we first consider uplink
MMSE receiver. We further propose a suboptimal MMSE receiver called \emph{partial MMSE}
and demonstrate that it
has virtually optimal performance.
In \cite{truong2013viability}, the authors study uplink performance of cellular mMIMO systems with distributed antenna clusters in each cell. The authors consider MMSE and MF receivers with coordination among distributed antenna clusters in each cell. 
In contrast we assume all distributed APs coordinate with each other to form the postcoding vectors and detect the signals transmitted by users.
In \cite{Hoydis} random matrix theory results are used to study performance of cellular mMIMO systems.
Motivated by \cite{Hoydis} we applied random matrix theory for deriving a tight approximation of the partial MMSE in cell-free systems as a function of large scale fading coefficients with cooperation among distributed APs. Since partial MMSE has almost optimal performance, our approximation is also very accurate for the optimal MMSE
receiver. Numerical results indicate that the obtained approximation is accurate even for a small number of APs and users.

In \cite{ashikhmin2012pilot} and \cite{adhikary2014}, LSFD (also known as pilot contamination postcoding) was proposed for interference reduction in cellular mMIMO systems. In LSFD  base stations cooperate, but only using large scale fading coefficients.
In this work we propose generalization of the LSFD receiver for cell-free mMIMO systems and derive
the SINR expression for it as a function of large scale fading coefficients. 

We further derive an expression for SINR of cell-free systems with MF receiver in the regime when the number of users is constant and the number of APs grows without a limit. Our result  shows that in this regime the system performance is limited by the coherent interference resulting from two or more users sharing the same pilot sequence.

In numerical experiments we evaluate the system performance under independent and correlated shadow fading models.
Results show that MMSE and LSFD receivers provide significant gain over MF receiver. MMSE receiver outperforms LSFD receiver while the latter has smaller complexity.

The paper is organized as follows. In section \ref{section:system_model} the system model and channel estimation are described. In section \ref{section:uplink_data}, we investigate MMSE, partial MMSE, and LSFD receivers in uplink transmission. Finally, numerical results are presented in section \ref{section:simulations}.

Throughout the paper  $\text{diag}\left(a_i\right)_{1\le i\le n}$ denotes diagonal matrix with $a_1,\cdots,a_n$ on its diagonal. 
If $\mathcal{S}_1=\left\{\alpha_1,\cdots,\alpha_n\right\}\in\mathbb{N}^{n\times 1}$ and $\mathcal{S}_2=\left\{\sigma_1,\cdots,\sigma_m\right\}\in\mathbb{N}^{m\times 1}$, where $\alpha_i$ and $\sigma_i$s are in the increasing order, then $\left[v_i\right]_{i\in\mathcal{S}_1}$ denotes the column vector $\left[v_{\alpha_1},\cdots,v_{\alpha_n}\right]^T$; and $\left[\left[v_{ij}\right]\right]_{i\in\mathcal{S}_1,j\in\mathcal{S}_2}$ denotes the $n\times m$ matrix $\left[\begin{smallmatrix}v_{\alpha_{1}\sigma_1},&\cdots,&v_{\alpha_{1}\sigma_m}\\ \vphantom{\int\limits^x}\smash{\footnotesize{\vdots}} &\vphantom{\int\limits^x}\smash{\footnotesize{\ddots}} &\vphantom{\int\limits^x}\smash{\footnotesize{\vdots}}\\ v_{\alpha_{n}\sigma_1},&\cdots,&v_{\alpha_{n}\sigma_m}\end{smallmatrix}\right]^T$.

\section{System Model and Channel Estimation} \label{section:system_model}

We consider a geographical area with $M$ randomly distributed single-antenna APs and $K$ single antenna users, assuming
that $K\ll M$. All APs are
connected to a network controller (NC) via an unspecified backhaul network.  All APs and users are perfectly synchronized in time.
The channel coefficient between AP $m$ and user $k$ is given by
\begin{align*}
g_{mk}=\sqrt{\beta_{mk}}h_{mk},
\end{align*}
where $\beta_{mk}$ is the large scale fading coefficient which accounts for path loss and shadow fading and $h_{mk}\sim\mathcal{CN}\left(0,1\right)$ is the small scale fading coefficient. The large scale fading coefficients change slowly over time and assumed to be known at the NC. The small scale fading coefficients are i.i.d. random variables that stay constant over a channel coherence interval.

We assume time-division duplex (TDD) protocol, i.e., all users synchronously send randomly assigned orthonormal pilot sequences ($\boldsymbol{\psi}_1,\cdots,\boldsymbol{\psi}_{\tau}\in\mathbb{C}^{\tau\times 1}$, where $\boldsymbol{\psi}_i^H\boldsymbol{\psi}_j=\delta(i-j)$) to allow APs to estimate channel coefficients, which they further send to the NC.

We consider short coherence interval (due to high user mobility) and therefore $\tau$ is small and $K>\tau$. Hence each pilot is reused by several users, which results in the pilot contamination, \cite{ashikhmin2012pilot},\cite{adhikary2014}.
In \cite{Hien}, a greedy pilot assignment scheme in cell-free systems has been introduced which is shown to improve performance of cell-free system compared with the random pilot assignment scheme. However, for simplicity we consider the random pilot assignment in the cell-free systems.

All users are partitioned into $\tau$ sets $\mathcal{S}_1,\cdots,\mathcal{S}_{\tau}$ in a way that users in $\mathcal{S}_j$ use pilot $\boldsymbol{\psi}_j$. Let $b_i$ be the index of the pilot sequence transmitted by th $i$th user. The received signal in the first step of the TDD protocol  at the $m$th AP is
\begin{align*}
	\boldsymbol{y}_m=\sqrt{\rho\tau}\sum_{i=1}^{K}g_{mi}\boldsymbol{\psi}_{b_i}+\boldsymbol{v}_m,
\end{align*}
where $\rho$ is the uplink transmit power of each user and $\boldsymbol{v}_m\in\mathbb{C}^{\tau\times1}\sim\mathcal{CN}(0,1)$ is additive Gaussian noise. AP $m$ computes the MMSE estimate of $g_{mk}$ as
\begin{align*}
	 \hat{g}_{mk}=\frac{\sqrt{\rho\tau}\beta_{mk}}{1+\rho\tau\sum_{i\in\mathcal{S}_{b_k}}\beta_{mi}}\boldsymbol{\psi}_{b_k}^H\boldsymbol{y}_m.
\end{align*}
It can be verified that $\hat{g}_{mk}$ and the channel estimation error $\tilde{g}_{mk}=g_{mk}-\hat{g}_{mk}$ are uncorrelated Gaussian random variables with distributions 
\begin{align*}
\hat{g}_{mk}\sim\mathcal{CN}\left(0,\alpha_{mk}\right),\ \ \
\tilde{g}_{mk}\sim\mathcal{CN}\left(0,\beta_{mk}-\alpha_{mk}\right),
\end{align*}
where $\alpha_{mk}=\frac{{\rho\tau}\beta_{mk}^2}{1+\rho\tau\sum_{i\in\mathcal{S}_{b_k}}\beta_{mi}}$.
Note that $\hat{g}_{mi}=\frac{\beta_{mi}}{\beta_{mk}}\hat{g}_{mk}$ for every $i,k\in\mathcal{S}_{b_k}$. Therefore, it is enough for AP $m$ to choose one user $u_{j}\in\mathcal{S}_{j}$ and send only the channel estimates $\hat{g}_{mu_{j}},~j=1,\cdots,\tau$ to the NC.

Let $\eta_i$ denote the power coefficient used by the $i$th user to transmit uplink data. For notation convenience we define
\begin{alignat*}{3}
A_i&\triangleq\text{diag}\left(\alpha_{mi}\right)_{1\le m\le M},\ \ \  &&B_i\triangleq\text{diag}\left(\beta_{mi}\right)_{1\le m\le M},\\
C_{i}&\triangleq B_i-A_i,  &&D_{\textcolor{white}{i}}\triangleq\rho\sum_{i=1}^{K}\eta_{i}C_{i}+I.
\end{alignat*}

\section{Uplink Data Transmission}\label{section:uplink_data}

At the second step of the TDD protocol, users send data symbols and the $m$th AP receives
\begin{align*}
y_m=\sqrt{\rho}\sum_{i=1}^K\sqrt{\eta_i}g_{mi}s_i+v_m,
\end{align*}
where $v_m\sim\mathcal{CN}(0,\sigma_z^2)$ is additive noise and $s_i$ is the data signal transmitted by the $i$th user. The NC uses estimates $\hat{g}_{mk}$
to form postcoding vectors ${\bf v}_k$ and obtains estimates of data signals $\hat{s}_k={\bf v}_k^H\left[y_1,\cdots,y_M\right]^T,~k=1,\cdots,K$. Using the worst-case uncorrelated additive noise, the uplink achievable rate of the $k$th user is $R=\mathbb{E}\left(\log_2\left(1+\text{SINR}_k\right)\right)$, with
\begin{align}
\text{SINR}_{k}\left({\bf v}_{k}\right)&=\frac{\rho\eta_{k}{\bf v}_{k}^H\hat{\boldsymbol{g}}_{k}\hat{\boldsymbol{g}}_{k}^H
{\bf v}_{k}}{{\bf v}_{k}^H\left(\rho\sum_{i\neq k}^K\eta_{i}
\hat{\boldsymbol{g}}_{i}\hat{\boldsymbol{g}}_{i}^H+D\right){\bf v}_{k}},\label{eq:SINR}
\end{align}
where $\hat{\boldsymbol{g}}_i=\left[\hat{g}_{1i},\cdots,\hat{g}_{Mi}\right]^T$. Note that achievable SINR of the $k$th user in (\ref{eq:SINR}) is obtained by taking into account the channel estimation error and pilot contamination effect.

\subsection{MMSE Receiver}

First, we consider MMSE receiver, which maximizes SINR of each user. The MMSE vector of the $k$th user is given by
\begin{align}
{\bf v}_k^{\text{\tiny{MMSE}}}=\sqrt{\rho\eta_k}\bigg(\rho\sum_{i=1}^K\eta_{i}\hat{\boldsymbol{g}}_i\hat{\boldsymbol{g}}_i^H+D\bigg)^{-1}\hat{\boldsymbol{g}}_k.\label{eq:opt}
\end{align}
Note that the MMSE vector in (\ref{eq:opt}) contains channel estimates of all users in the network. Thus, it is optimal in the sense that it maximizes SINR of each user. Whereas in cellular systems, the MMSE vector at cell $\ell$ only contains channel vectors of  cell $\ell$ and the second-order statistics of the channel coefficients between base station at cell $\ell$ and all users in the network \cite{truong2013viability,Hoydis}. Achievable SINR of the $k$th user with MMSE receiver is given by
\begin{align}
\text{SINR}_{k}^\text{\tiny{MMSE}}&=\text{SINR}_{k}\left( {\bf v}_k^{\text{\tiny{MMSE}}}   \right)\nonumber\\
&=\frac{\hat{\boldsymbol{g}_{k}}^H
\left(\rho\sum_{i=1}^K\eta_{i}\hat{\boldsymbol{g}}_i\hat{\boldsymbol{g}}_i^H+D\right)^{-1}\hat{\boldsymbol{g}_{k}}}{\frac{1}{\rho\eta_{k}}-\hat{\boldsymbol{g}_{k}}^H
\left(\rho\sum_{i=1}^K\eta_{i}\hat{\boldsymbol{g}}_i\hat{\boldsymbol{g}}_i^H+D\right)^{-1}\hat{\boldsymbol{g}_{k}}}.
\end{align}

The Monte Carlo simulation of $R^{\text{\tiny{MMSE}}}_k=\log_2\left(1+\text{SINR}_k^{\text{\tiny{MMSE}}}\right)$ requires long averaging over small scale fading coefficients $h_{mk}$. Hence it is desirable to have an approximation of $R^{\text{\tiny{MMSE}}}_k$  as a function of large scale fading coefficients only. 
The correlation between the channel estimates (i.e., $\hat{g}_{mi}=\frac{\beta_{mi}}{\beta_{mk}}\hat{g}_{mk}$ for $i,k\in\mathcal{S}_{b_k}$) does not allow us to use random matrix theory tools (\cite[Theorem 1,2]{wagner},\cite{Hoydis}) to achieve this goal. Below we propose a \emph{partial MMSE} receiver whose performance is very close to the performance of the MMSE receiver and allows us to overcome this problem.

\subsection{Partial MMSE Receiver}

 Let $\mathcal{I}_k=\mathcal{S}_{b_k}\cup\big\{u_{1}^{(k)},\cdots,u_{\tau}^{(k)}\big\}$, where $u_j^{(k)}\in\mathcal{S}_j$ is the index of a user from $\mathcal{S}_j$ whose selection rule is discussed later.
The partial MMSE vector for user $k$ is then defined by 
\begin{align}
{\bf v}_k^{\text{\tiny{PMMSE}}}&=\sqrt{\rho\eta_k}\bigg(\hspace{-0.1em}\rho\hspace{-0.3em}\sum_{i\in \mathcal{I}_k}\eta_i\hat{\boldsymbol{g}}_i\hat{\boldsymbol{g}}_i^H\hspace{-0.2em}+\hspace{-0.2em}\rho\hspace{-0.3em}\sum_{i\notin \mathcal{I}_k}\mathbb{E}\big(\eta_i\hat{\boldsymbol{g}}_i\hat{\boldsymbol{g}}_i^H\big)\hspace{-0.2em}+\hspace{-0.2em}D\hspace{-0.1em}\bigg)^{-1}\hspace{-0.3em}\hat{\boldsymbol{g}}_k\nonumber\\
&= \sqrt{\rho\eta_k}\bigg(\hspace{-0.1em}\rho\hspace{-0.3em}\sum_{i\in \mathcal{I}_k}\eta_i\hat{\boldsymbol{g}}_i\hat{\boldsymbol{g}}_i^H\hspace{-0.2em}+\hspace{-0.2em}Q\hspace{-0.1em}\bigg)^{-1}\hspace{-0.3em}\hat{\boldsymbol{g}}_k,\label{eq:subopt}
\end{align}
where 
$$Q=\rho\hspace{-0.2em}\sum_{i\notin\mathcal{I}_k}\hspace{-0.2em}\eta_iB_{i}\hspace{-0.2em}+\hspace{-0.2em}\rho\hspace{-0.2em}\sum_{i\in\mathcal{I}_k}\hspace{-0.2em}\eta_iC_{i}\hspace{-0.2em}+\hspace{-0.2em}I.$$
 Note that $\mathcal{I}_k$ contains all users that cause coherence interference to user $k$ and one user from each non-coherent interference group $\mathcal{S}_j,~j\neq k$. 
Note that in mMIMO systems, the coherent interference is the dominant impairment which limits the system performance when number of antennas increase without bound. Therefore, in the partial MMSE receiver we include channel vectors of all users that use the same pilot sequence as user $k$. 
The users $u_{1}^{(k)},\cdots,u_{\tau}^{(k)}$ should be chosen such that vectors $\hat{\boldsymbol{g}}_i,i\in\mathcal{I}_k$ in~(\ref{eq:subopt}) have the major contribution in~(\ref{eq:opt}) and hence (\ref{eq:subopt}) becomes close to (\ref{eq:opt}). Numerical results show that a random selection of users $u_{1}^{(k)},\cdots,u_{\tau}^{(k)}$ from the corresponding sets $\mathcal{S}_1,\cdots,\mathcal{S}_{\tau}$ leads to poor performance (see Figure \ref{fig:1}). A method for smart choice of these users can be formulated as following
\begin{align}
u_j^{(k)}=\text{arg }\underset{i\in\mathcal{S}_j}{\text{max}}\ \boldsymbol{\beta}_{k}^T\boldsymbol{\beta}_{i}, \ \ \ j=1,\cdots,\tau,\label{eq:Choose}
\end{align}
where $\boldsymbol{\beta}_{i}=\left[\beta_{1i},\cdots,\beta_{Mi}\right]^T$. In other words, we choose user $u_j^{(k)}\in\mathcal{S}_j$ that is in the close vicinity of the $k$th user.
The $\text{SINR}_{k}^{\text{\tiny{PMMSE}}}$ can be obtained by substituting ${\bf v}_{k}^{\text{\tiny{PMMSE}}}$ in (\ref{eq:SINR}).


In the following theorem we apply random matrix theory to obtain an asymptotic approximation 
of $R^{\text{\tiny{PMMSE}}}_k=\log_2\left(1+\text{SINR}^{\text{\tiny{PMMSE}}}_k\right)$ when $M$ and $K$ grow infinitely large while the ratio $\sfrac{M}{K}$ is finite. This asymptotic result is used as an approximation for finite values of $M$ and $K$ similar to \cite{wagner} and \cite{Hoydis} in which the approximations are derived for MISO broadcast channel and cellular systems respectively.

\begin{Theorem}\label{theorem:MMSE_Asymp}
Assume matrices $A_{i}$, $C_{i}~i=1,\cdots,K$ have uniformly bounded spectral norms.
For the partial \textnormal{MMSE} receiver defined in (\ref{eq:subopt}), when $M$ and $K$ grow large such that $0<\liminf_M\frac{M}{K}\le\limsup_M\frac{M}{K}<\infty$,  we have 
\begin{align*}
\textnormal{R}_{k}^{\textnormal{\tiny{PMMSE}}}-\log_2\left(1+\fixwidehat{\textnormal{SINR}}_{k}^{\textnormal{\tiny{PMMSE}}}\right)\overset{\textnormal{a.s.}}{\underset{M,K\to\infty}{\xrightarrow{\hspace*{0.7cm}}}}0,
\end{align*}
\begin{figure*}[!t]
\normalsize
\begin{align}
\fixwidehat{\textnormal{SINR}}_{k}^{\text{\tiny{PMMSE}}}=\frac{\rho\eta_k\lambda_{k}^2}{\theta\left(D\right)+
\rho\sum\limits_{i\in \mathcal{S}_{b_k}\setminus \{k\}}\eta_i\lambda_i^2+\rho\sum\limits_{i\notin\mathcal{I}_k}\eta_i\theta\left(A_{i}\right)
+\rho\sum\limits_{\substack{i\in\mathcal{I}_k\setminus \mathcal{S}_{b_k}}}\frac{\eta_i\theta\left(A_{i}\right)}{\left(1+\rho\frac{\eta_i}{M}\textnormal{tr}\left(A_{i}T''_i\right)\right)^2}}\label{eq:SINR_subopt_asymp}
\end{align}
\hrulefill
\vspace*{4pt}
\end{figure*}
where $\fixwidehat{\textnormal{SINR}}_{k}^{\textnormal{\tiny{PMMSE}}}$ is defined in (\ref{eq:SINR_subopt_asymp}) and all parameters in $\fixwidehat{\textnormal{SINR}}_{k}^{\textnormal{\tiny{PMMSE}}}$ are summarized in Table \ref{table:parameters} \footnote{Generalized matrix inversion lemma and \cite[Theorem 1,2]{Hoydis} are used to derive the asymptotic approximation. Due to lack of space, derivations are skipped. }.

\begin{table}
\centering
\renewcommand{\arraystretch}{1.9}
\setlength{\tabcolsep}{4pt}

\caption{Parameter definitions in Theorem \ref{theorem:MMSE_Asymp}.}

\begin{tabular}{| c|| c |  }\hline
\rule{0pt}{4ex}
$\delta_i^{(t)}$ & $\frac{\rho\eta_i}{M}\textnormal{tr}\,A_{i}\bigg(\frac{\rho}{M}\sum\limits_{j\in\mathcal{I}_k\setminus\mathcal{S}_{b_k}}\frac{\eta_jA_{j}}{1+\delta_j^{(t-1)}}+\frac{1}{M}Q\bigg)^{-1}$\\ \hline

\rule{0pt}{3ex}

$\delta_j$ & $\underset{t\to\infty}{\lim}\delta_j^{(t)}$, \textnormal{with} $\delta_j^{(0)}=M$\\ \hline

\rule{0pt}{4ex}

$T$ &  $\bigg(\frac{\rho}{M}\sum\limits_{j\in\mathcal{I}_k\setminus\mathcal{S}_{b_k}}\frac{\eta_jA_{j}}{1+\delta_j}+\frac{1}{M}Q\bigg)^{-1}$ \\ \hline

\rule{0pt}{4ex}

$[J]_{jl}$ & $\dfrac{\frac{\rho^2}{M}\textnormal{tr}\left(\eta_j\eta_lA_{j}TA_{{l}}T\right)}{M\left(1+\delta_{l}\right)^2}, \ \ \ j,l\in\mathcal{I}_k\setminus\mathcal{S}_{b_k}$\\ \hline

$\boldsymbol{\delta}'$ & $\big[\delta'_j\big]_{j\in\mathcal{I}_k\setminus\mathcal{S}_{b_k}}=\left(I-J\right)^{-1}\left[\frac{\rho\eta_j}{M}\textnormal{tr}\left(A_{j}THT\right)\right]_{j\in\mathcal{I}_k\setminus\mathcal{S}_{b_k}}$\\ \hline

\rule{0pt}{3.5ex}

$T'\left(H\right)$ & $THT+T\frac{\rho}{M}\hspace{-0.4em}\sum\limits_{j\in\mathcal{I}_k\setminus\mathcal{S}_{b_k}}\hspace{-0.3em}\frac{\eta_jA_{j}\delta_j'}{\left(1+\delta_j\right)^2}T$\\  \hline

\rule{0pt}{4ex}

${\delta''}_{i}^{(t)}$  &$\frac{\rho\eta_i}{M}\textnormal{tr}\,A_{i}\bigg(\frac{\rho}{M}\sum\limits_{\substack{j\in\mathcal{I}_k\setminus\{n\}}}\frac{\eta_jA_j}{1+{\delta''}_{j}^{(t-1)}}+\frac{1}{M}Q\bigg)^{-1}$\\ \hline

$\delta''_j$ & $\underset{t\to\infty}{\lim}{\delta''_j}^{(t)}$, \textnormal{with} ${\delta''_j}^{(0)}=M$\\ \hline

\rule{0pt}{4ex}

$T''_i$ & $\bigg(\frac{\rho}{M}\sum\limits_{\substack{j\in\mathcal{I}_k\setminus \{i\}}}\frac{\eta_jA_j}{1+\delta''_{j}}+\frac{1}{M}Q\bigg)^{-1}$\\ \hline

\rule{0pt}{3ex}

$\boldsymbol{\gamma}_{i}$ &  $\frac{\sqrt{\rho}}{M}\left[\sqrt{\eta}_j\textnormal{tr}\left(A_{i}^{1/2}A_{j}^{1/2}T\right)\right]_{j\in\mathcal{S}_{b_i}}$ \\ \hline

$\Gamma$ &  $I+\frac{\rho}{M}\left[\left[\sqrt{\eta_i\eta_j}\text{tr}\left(A_i^{1/2}A_j^{1/2}T\right)\right]\right]_{i\in\mathcal{S}_{b_k},j\in\mathcal{S}_{b_k}}$\\ \hline

\rule{0pt}{3ex}

$\boldsymbol{\nu}_{i}\left(H\right)$  & $\frac{\sqrt{\rho}}{M^2}\left[\sqrt{\eta_j}\textnormal{tr}\left(A_{j}^{1/2}A_{i}^{1/2}T'\left(H\right)\right)\right]_{j\in\mathcal{S}_{b_k}}$\\ \hline

$N\left(H\right)$ & $\frac{\rho}{M^2}\left[\left[\sqrt{\eta_i\eta_j}\text{tr}\left(A_i^{1/2}A_j^{1/2}T'\left(H\right)\right)\right]\right]_{i\in\mathcal{S}_{b_k},j\in\mathcal{S}_{b_k}}$\\ \hline

\rule{0pt}{3ex}

$\lambda_{i}$ & $\frac{1}{M}\textnormal{tr}\left(A_{i}^{1/2}A_{k}^{1/2}T\right)
-\boldsymbol{\gamma}_{k}^T\Gamma^{-1}\boldsymbol{\gamma}_{i}$\\ \hline

$\theta\left(H\right)$  &$\!\begin{aligned}&\tfrac{1}{M^2}\textnormal{tr}\left(A_{k}T^{\prime}\left(H\right)\right)-2\textnormal{Re}\left(\boldsymbol{\nu}_{k}
\left(H\right)^T\Gamma^{-1}\boldsymbol{\gamma}_{k}\right)\\
&+\boldsymbol{\gamma}_{k}^T\Gamma^{-1}N
\left(H\right)\Gamma^{-1}\boldsymbol{\gamma}_{k}\end{aligned}$\\ \hline

\end{tabular}

\label{table:parameters}
\end{table}

\end{Theorem}


Note that the approximation $\fixwidehat{\textnormal{SINR}}_{k}^{\textnormal{\tiny{PMMSE}}}$ in (\ref{eq:SINR_subopt_asymp}) is a function of large scale fading coefficients \emph{only}, and though it has a long formulation, it can be easily calculated numerically for large values of $M$ and $K$.

\subsection{Large Scale Fading Decoding}
Next, we propose the LSFD receiver for cell-free systems. The main idea of LSFD is that only large scale fading coefficients are transmitted to NC from APs. Since these coefficients are independent of frequency and change (about $40$ times) slower than small scale fading coefficients, LSFD allows one to reduce the backhaul traffic, which can be very desirable in real life systems. 

By using matched filter, the $m$th AP computes $\tilde{s}_{mk}=\hat{g}_{mk}^*y_m$ for one user $k\in\mathcal{S}_{b_k}$, and sends them to the NC.
The NC computes $\tilde{s}_{mi}=\frac{\beta_{mi}}{\beta_{mk}}\tilde{s}_{mk},~i\in\mathcal{S}_{b_k}$ and estimates data symbol $s_k$ by using linear combination of all received signals as following
\begin{align}
\hat{s}_{k}=\sum_{m=1}^M\sum_{i=1}^K\text{v}^{*}_{mki}\tilde{s}_{mi}.\label{eq:s_hat}
\end{align}
The NC computes postcoding coefficients $\text{v}_{kmi}$ and power coefficients $\eta_k$ as a function of large scale fading coefficients only. 
\begin{Lemma} \label{lemma:s_hat}
The estimate of data symbol $\hat{s}_k$ in (\ref{eq:s_hat}) can be simplified as
\begin{align}
\hat{s}_{k}={\bf v}_k^H\tilde{\boldsymbol{s}}_k,\label{eq:s_hat2}
\end{align}
where ${\bf v}_k=\left[\text{v}_{1k},\cdots,\text{v}_{Mk}\right]^T$ and $\tilde{\boldsymbol{s}}=\left[\tilde{s}_{1k},\cdots,\tilde{s}_{Mk}\right]^T$.
\end{Lemma} 
The proof of Lemma \ref{lemma:s_hat} follows directly from the fact that assignment $\text{v}_{mki}=0,i\not\in \mathcal{S}_{b_k}$, in (\ref{eq:s_hat}) does not result in any performance loss, and $\tilde{s}_{mi}=\frac{\beta_{mi}}{\beta_{mk}}\tilde{s}_{mk},~i\in\mathcal{S}_{mi}$.

\begin{Theorem}\label{theorem:Rate} Achievable SINR of the $k$th user with LSFD receiver is given by
\begin{align*}
\textnormal{SINR}_{k}\left({\bf v}_k\right)=\frac{\rho\eta_{k}{\bf v}_{k}^H\boldsymbol{\mu}_{k}\boldsymbol{\mu}_{k}^H{\bf v}_{k}}{\rho\sum_{\substack{i\in\mathcal{S}_{b_k}\setminus \{k\}}}\eta_{i}{\bf v}_{k}^H\boldsymbol{\mu}_{i}\boldsymbol{\mu}_{i}^{H}{\bf v}_{k}+{\bf v}_{k}^H\Lambda{\bf v}_{k}},
\end{align*}
where $\Lambda=\textnormal{diag}\left(\rho\sum_{i=1}^K\eta_i\alpha_{mk}\beta_{mi}+\alpha_{mk}\right)_{1\le m\le M}$ and $\mu_i=\left[\frac{\rho\tau\beta_{mk}\beta_{mi}}{1+\rho\tau\sum_{j\in\mathcal{S}_{b_{i}}}\beta_{mj}}\right]_{1\le m\le M}$.
\end{Theorem}

We can show that the optimal ${\bf v}_{k}^{\text{\tiny{LSFD}}}$ which maximizes SINR of each user is given by
\begin{align*}
{\bf v}^{\text{\tiny{LSFD}}}_{k}=\left(\rho\sum_{i\in\mathcal{S}_{b_k}\setminus \{k\}}\eta_{i}\boldsymbol{\mu}_{i}\boldsymbol{\mu}_{i}^{H}+\Lambda\right)^{-1}\boldsymbol{\mu}_{k}.
\end{align*}
The associated SINR of the $k$th user is given by
$$\text{SINR}_k^{\text{\tiny{LSFD}}}=\rho\eta_k\boldsymbol{\mu}_k^H\left(\rho\sum_{i\in\mathcal{S}_{b_k}\setminus \{k\}}\eta_{i}\boldsymbol{\mu}_{i}\boldsymbol{\mu}_{i}^{H}+\Lambda\right)^{-1}\boldsymbol{\mu}_{k}.$$
To obtain power coefficients one can apply max-min power allocation problem with per user transmit power constraints as
following
\begin{subequations}
\label{eq:maxmin}
\begin{align}
\max_{\boldsymbol{\eta}} \min_{k} \ &R_k^{\text{\tiny{LSFD}}}=\log_2\left(1+\text{SINR}^{\text{\tiny{LSFD}}}_{k}\right),\label{eq:maxmin_SINR}\\
\text{s.t. }\ &\eta_{i}\le 1,\ \ \ \ i=1,\cdots,K.\label{eq:maxmin_constraints}
\end{align}
\end{subequations}
\begin{Lemma}
The objective function $\min_{k} \ R_k^{\textnormal{\tiny{LSFD}}}\left(\boldsymbol{\eta}\right)$ in (\ref{eq:maxmin_SINR}) is a quasiconcave function of $\boldsymbol{\eta}=\left[\eta_1,\cdots,\eta_K\right]^T$ and constraints (\ref{eq:maxmin_constraints}) are convex.
\end{Lemma}
Since the power allocation problem (\ref{eq:maxmin}) is quasiconcave, bisection method \cite[Chapter~4.2.5]{Boyd} can be used to solve it.

We wrap up this section by providing the SINR expression for MF receiver when the number of APs grows without limit.
\begin{Theorem}\label{theorem:SINR_LSFD}
Achievable SINR of the $k$th user for MF receiver, i.e., ${\bf v}_k^{\textnormal{\tiny{MF}}}=\left[1,\cdots,M\right]^T$,  with unlimited number of APs ($M\to\infty$ and $K=\text{constant}$) and independent large scale fading coefficients is given by
\begin{align}
\textnormal{SINR}_{k}\left({\bf v}_k^{\textnormal{\tiny{MF}}}\right)\overset{\textnormal{a.s.}}{\underset{M\to\infty}{\xrightarrow{\hspace*{0.6cm}}}}\frac{\eta_k\big(\mathbb{E}\left(\beta_{mk}c_{mk}\right)\big)^2}{\sum\limits_{i\in\mathcal{S}_{b_k}\setminus\{k\}}\eta_i\big(\mathbb{E}\left(\beta_{mk}c_{mi}\right)\big)^2},\label{eq:SINR_MF}
\end{align}
where $c_{mi}=\frac{\rho\tau\beta_{mi}}{1+\rho\tau\sum_{j\in\mathcal{S}_{b_i}}\beta_{mj}}$ and the expected value is over location of APs (index $m$).
\end{Theorem}
Note that the denominator in (\ref{eq:SINR_MF}) corresponds to power of the pilot contamination related interference. Similar to the cellular mMIMO systems, SINR of the $k$th user using MF receiver is limited by the effect of pilot contamination. However,  unlink cellular systems, in which SINR depends on the large scale fading coefficients, SINR of cell-free system is a constant quantity in the limit of an infinite number of APs.


\section{Numerical Results} \label{section:simulations}
We consider a square dense urban area of $2\times 2$ km$^2$ with $M$ randomly located APs and $K$ randomly located users. The area is  wrapped around to avoid boundary effects. For large scale fading coefficients we consider a three-slope path loss model \cite{Path} as follows
\begin{align}
\beta_{mk} &= \renewcommand{\arraystretch}{1.45}\left\{\begin{array}{@{}l@{\quad}l@{}}
c_0&\quad d_k\le 0.01 \text{ km}\\
\dfrac{c_1}{d_{mk}^{2}}&\quad 0.01 \text{km}<d_k\le 0.05 \text{ km}\\
\dfrac{c_2z_{mk}}{d_{mk}^{3.5}}&\quad d_k> 0.05 \text{ km}
 \end{array}\right.\kern-\nulldelimiterspace,\label{eq:large_scale_fading}
\end{align}
where $d_{mk}$ is the distance in kilometers between user $k$ and the AP $m$, and $z_{mk}$ is the log-normal shadow fading, i.e., $10\log_{10}z_{mk}\sim\mathcal{N}(0,\sigma_{\text{shad}}^2)$ with $\sigma_{\text{shad}}=8$ dB. For $d_k>0.05$ km we use COST-231 Hata propagation model
\begin{align*}
10\log_{10}c_2=&-46.3 -33.9\log_{10}f+13.82\log_{10}h_B\\
&+ (1.1\log_{10}f-0.7)h_R-(1.56\log_{10}f-0.8),
\end{align*}
where $f = 1900$ MHz is the carrier frequency, $h_B = 15$ m is
the AP antenna hight, and $h_R = 1.65$ m is the user
antenna hight. Parameters $c_1$ and $c_2$ in (\ref{eq:large_scale_fading}) are chosen in the way that path loss remains continuous at boundary points.

To model the correlation between large scale fading coefficients caused by closely located users and/or APs, we use the correlation model from \cite{Hien} with $\delta=0.5$ and $d_{\text{decorr}}=0.1$ km. The noise variance is $\sigma_v^2=290\times\kappa\times B\times NF$, where $\kappa$, $B$, and $NF$ are Boltzmann constant, bandwidth ($20$ MHz) and noise figure ($9$ dB) respectively. We assume users transmit with equal power $\eta_i=1,~i=1,\cdots,K$ and $\rho=200$ mW.

\begin{figure}[t]
	\centering
	\includegraphics[trim=0.96cm 0.05cm 6.52cm 17.94cm, clip=true,width=0.5\textwidth]{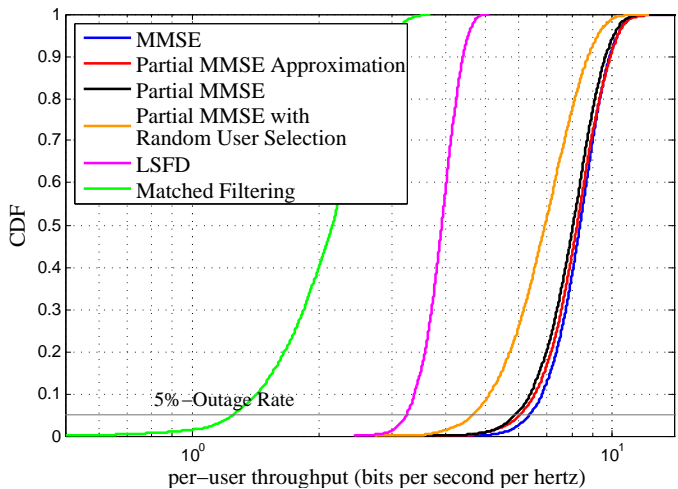}
	\caption{CDFs of the achievable per-user rates for LSFD and MMSE receivers with $M=1000$, $K=50$, and $\tau=10$.}
	\label{fig:1}
\end{figure}

\begin{figure}[t]
	\centering
	\includegraphics[trim=0.93cm 0.15cm 7.05cm 17.94cm, clip=true,width=0.5\textwidth]{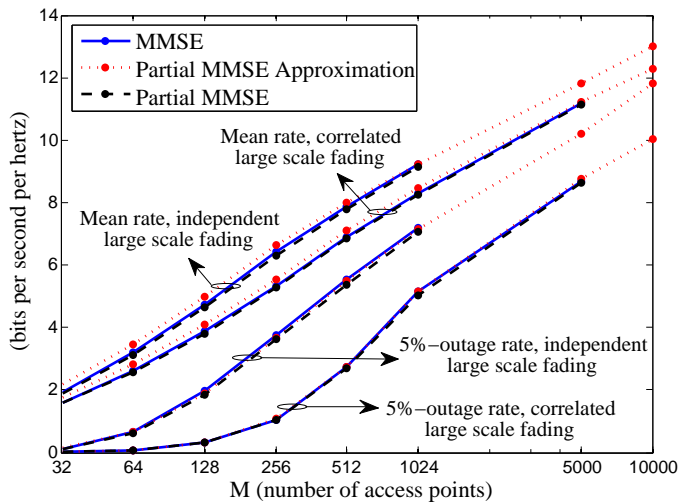}
	\caption{$5\%$-Outage and mean rates versus $M$ for correlated and independent large scale fading with $K=16$ and $\tau=4$.}
	\label{fig:2}
\end{figure}

\begin{figure}[t]
	\centering
	\includegraphics[trim=0.81cm 0.04cm 7.5cm 17.94cm, clip=true,width=0.5\textwidth]{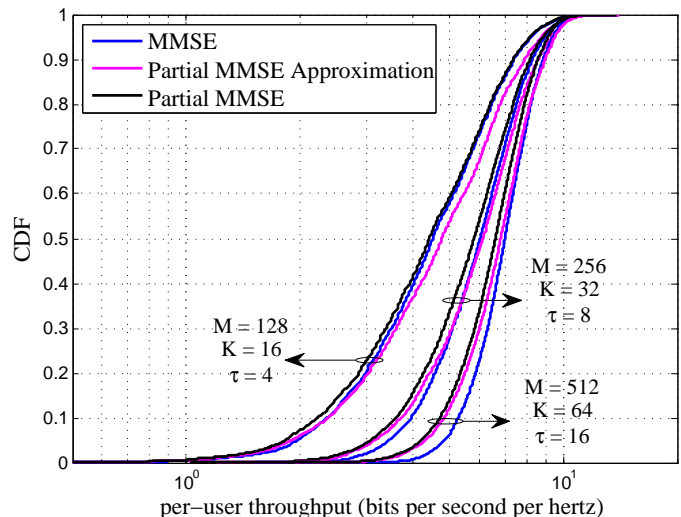}
	\caption{CDFs of the achievable per-user rates for MMSE receivers with different number of APs and users.}
	\label{fig:3}
\end{figure}

In figure \ref{fig:1}, CDFs of $R^{\text{\tiny{MMSE}}}$, $R^{\text{\tiny{PMMSE}}}$ with heuristic approach given in (\ref{eq:Choose}), $R^{\text{\tiny{PMMSE}}}$ with random user selection, $\hat{R}^{\text{\tiny{PMMSE}}}=\log_2\left(1+\fixwidehat{\text{SINR}}^{\text{\tiny{PMMSE}}}\right)$, and $R^{\text{\tiny{LSFD}}}$ with independent large scale fading coefficients are presented. The CDF of per-user throughput achieved by MF receiver \cite{Hien} is also included in the figure for comparison. The horizontal line corresponds to $5\%$-outage rates which represents the smallest rate among $95\%$ of the best users.
One can observe that the asymptotic approximation of MMSE receiver is very tight. MMSE and LSFD receivers provide respectively 5.1-fold and 2.6-fold gain over the MF receiver in terms of $5\%$-outage rate. Performance of the LSFD receiver lies between the simple MF receiver and MMSE receivers. Compared to the MMSE receiver, LSFD reduces the overall complexity of the system.  

Figure \ref{fig:2} shows $5\%$-outage and mean values of $R^{\text{\tiny{MMSE}}}$, $R^{\text{\tiny{PMMSE}}}$, $\hat{R}^{\text{\tiny{PMMSE}}}$ versus number of APs under independent and correlated shadow fading.
One can observe that in all considered scenarios the partial MMSE is virtually optimal and our approximation $\hat{R}_{\text{\tiny{MMSE}}}^{\text{\tiny{partial}}}$ is very accurate. The shadow fading correlation significantly affects the system performance.

The CDFs of per-user rates for different number of APs and users are plotted in Figure \ref{fig:3}. The ratio between APs and users is constant in all cases, i.e., $\sfrac{M}{K}=8$ and $\sfrac{K}{\tau}= 4$. We observe that the $5\%$-outage rate of MMSE and partial MMSE receivers increase as the network size increases.


\section{Conclusion}
In this paper we studied the uplink performance of cell-free systems with MMSE and LSFD receivers. A suboptimal MMSE receiver, which is more tractable to study the asymptotic behavior of the cell-free systems, is introduced. Rates achieved by MMSE, partial MMSE, and asymptotic approximation are very close. The asymptotic approximation is very accurate even for small number of APs and users. LSFD receiver in cell-free systems is introduced. LSFD receiver depends only on the large scale fading coefficients.
MMSE and LSFD receivers demonstrate significant gain over MF receiver. There is a considerable gap between MMSE and LSFD receivers.




\end{document}